\magnification=\magstep 1
\baselineskip = 15.4pt 
\parskip = \smallskipamount

\newcount\ftnumber
\def\ft#1{\global\advance\ftnumber by 1
          {\baselineskip=12pt \footnote{$^{\the\ftnumber}$}{#1 }}}
\def\section#1{\vskip 12pt \centerline{{\sl #1}} \vskip 6pt}
\font\title = cmr10 scaled 1440        
        
{\title
\centerline{QBism as CBism:}
\vskip 10pt
\centerline{Solving the Problem of ``the Now''}}
\vskip 20pt
\centerline{N. David Mermin}
\vskip 1pt
\centerline{Laboratory of Atomic and Solid State Physics}
\vskip 1pt
\centerline{Cornell University, Ithaca NY 14853}
\vskip 15pt
{\narrower\narrower

\noindent The QBist view of science, first put forth to dispel the fog from quantum foundations, also clears up a longstanding  puzzle in classical physics.

}

\vskip 10pt

\section{I. QBism and CBism}

In a  Physics Today Commentary,\ft{N. David Mermin, ``Fixing the Shifty split'', Physics Today, July 2012, pps.8-10.   Letters and my replies are in the December 2012 and June 2013 issues.} and more carefully, extensively, and convincingly with Chris Fuchs and Ruediger Schack,\ft{Christopher A. Fuchs, N. David Mermin, and Ruediger Schack, ``An Introduction to QBism
with an Application to the Locality of Quantum Mechanics'', arXiv:1311.5253.} I  argued that stubborn longstanding problems in the interpretation of quantum mechanics fade away if one takes literally  Niels Bohr's dictum\ft{ Niels Bohr,  {\it Atomic Theory and the Description of Nature\/},
Cambridge, 1934, p. 18.} that the purpose of science is not to reveal ``the real essence of the phenomena'' but to find ``relations between the manifold aspects of our experience.'' Here I note that the view of science as a tool that each of us can use to organize our own personal experience, called QBism\ft{``QBism" was initially an abbreviation for ``Quantum Bayesianism".   This can lead to distracting and irrelevant disagreements among different schools of Bayesianism.   That probabilities are personal judgments was put most forcibly by Bruno de Finetti, and if ``B'' has to stand for anything I would expand ``QBism" to ``Quantum Brunoism.''  Today, however, the term is familiar enough to stand on its own without any interpretation of the initials.}
 by Fuchs and Schack, clarifies more than just quantum foundational problems.   Recognizing that science is about the subject (the user\ft{QBists like to call the user an ``agent'', often named Alice.} of science) and not just about the object (the world external to that user)
can eliminate well entrenched confusion in  classical physics too. 
 
 When the QBist view of science is used  in a traditionally classical setting, I call it CBism\ft{Here C stands for ``Classical'' and B, for nothing at all, though it could  stand for the Bohr who wrote the dictum I cite above.       The name also brings to mind the racehorse Seabiscuit, who, after a slow start, had a spectacular career.}  to emphasize that the ability of QBism to solve conceptual puzzles unexpectedly extends to issues in which quantum physics plays no explicit role.   The double terminology is only to underline this fact, since QBism and CBism have identical understandings of science.  
 
In what follows I argue that CBism disposes of the  longstanding, vexing, and entirely classical problem of ``the Now''.   

\section{II. The Problem of the Now}
 
 Rudolf Carnap succinctly stated the problem of the Now in his report of a conversation with Albert Einstein:\ft{Rudolf Carnap, ÒIntellectual AutobiographyÓ in {\it The Philosophy of Rudolf Carnap\/}, ed. Paul A. Schilpp. Open Court Publishing Co., 1963, p.~37.}   
``Einstein said that the problem of the Now worried him seriously. He explained that the experience of the Now means something special for man, something essentially different from the past and the future, but that this important difference does not and cannot occur within physics. That this experience cannot be grasped by science seemed to him a matter of painful but inevitable resignation.''

The issue here is {\it not\/}  that the simultaneity of two different events in different places depends on frame of reference.    The issue  is that physics seems to have nothing whatever  to say about the local Now at a single event.\ft{By an ``event'' I mean an experience whose duration and location are restricted enough that for the purpose at hand it can usefully be represented as a point in space and time.} This apparent silence is a puzzle even without the relativity of simultaneity.   Physics, both pre- and post-relativistic, deals only with relations between one time and another.  Nevertheless a local present moment --- the Now --- is immediately evident as such to each and every one of us.  My experience of the Now is a primitive fact.   It simply can't be argued with.\ft{Actually anything can be argued with.   Last month I told an eminent theoretical physicist, who seemed to have shown a glimmer of sympathy for the QBist view of science, that I was working on an essay explaining how QBism solves the classical problem of the Now.   ``Ah,'' he said, ``you're going to explain why we all have  that illusion.''}  {\it Sum; ergo Nunc est.\/}   How can there be no place in physics for something as obvious as that?
 
My Now is a special event for me as it is happening.      The Now is distinguished from all the other events I have experienced by being the actual current state of affairs.  I can distinguish it from earlier events (former Nows) which I merely can remember, and from later events which I only can  anticipate or imagine. My  remembered past terminates in my Now.   The status of any particular event as my Now is fleeting, since it fades into a memory with the emergence of subsequent Nows.\ft{Philosophers speak of the problem of ``transience.''   I recommend  Steven Savitt:\vskip1pt
\centerline{http://faculty.arts.ubc.ca/ssavitt/Research/The\%20Transient\%20nows\%20(Rev).pdf.}}

Obvious as the human content of the preceding paragraph is to each and every one of us,\ft{It is not obvious to a distinguished philosopher of science, who recently had this to say about an unpublished, unarXived talk on the Now that I gave at the Perimeter Institute in 2009 [a video is at http://pirsa.org/09090077]: ``A distinguished quantum theorist insisted that the past is just a model we invent to make sense of present evidence and not to be taken literally.$\,\ldots$The time snobs' chauvinism of the present moment slides easily into solipsism.'' [Huw Price,  Science 341, 960-961,  30 August 2013.]  The QBist (CBist) recognition that the subject in science is as important as the object often elicits charges of solipsism, even though the multiplicity of subjects (agents) and their ability to communicate with each other is a crucial and explicit part of both the general QBist story and the particular CBist application I describe here, particularly in Section V below. 
}  such a Now is absent from the familiar physical description of space-time.    In physics all the events experienced by any person constitute a time-like curve, and there is nothing about any point to give it a special status as Now.  But my experience suggests that my world-line ought to terminate in something like a glowing point,  signifying my Now.   That glow should move in the direction of increasing time, as my world-line grows to accommodate more of my experience.   Yet there is nothing like this in the conventional physical description of my space-time trajectory.

The problem of the Now will not be solved by discovering new physics behind that  glowing point.  Nor is it solved by dismissing the Now as an ``illusion'' or as  ``chauvinism of the present moment.''   It is solved by identifying the mistakes that lead us to conclude, contrary to all our experience, that there is no place for the Now in our  physical description of the world.

\section{III.  The mistakes}

There are two mistakes.   The first is our  deeply ingrained unwillingness, noted above, to acknowledge that whenever anybody uses science it has a subject  as well as an object.
It is the well-established habit
of each of us to leave ourself --- the experiencing subject --- completely out of the story told by physics.\ft{QBism grew out of the efforts of Caves, Fuchs, and Schack to examine the implications for quantum foundations of interpreting probabilities as expressions of the personal judgment of an agent.  The recognition that science has a subject as well as an object (which, of course, is conducive to a personalist view of probability) provides a broader base for QBism.    The application here of CBism to the problem of the Now is thoroughly QBist,  even though probability plays no part in it.}$^,$\ft{Erwin Schr\"odinger returned repeatedly in his later years to deploring the exclusion of the perceiving subject from science, but I am not aware that he ever blamed this exclusion for the confusion at the foundations of quantum mechanics, or the apparently problematic nature of the Now.   See ``Nature and the Greeks'' (1954)   and ``Science and Humanism'' (1951), published together by Cambridge University Press, 1996; Section three and onward of ``Mind and Matter" (1958), published in ``What is Life'' by Cambridge, 1992; and ``What is Real?'' (1960),  reprinted in ``My View of the World'', Ox Bow Press, Woodbridge Connecticut, 1983.    Schr\"odinger did mention the importance  in quantum mechanics of the subject-object relation in a little-known  1931 letter to Arnold Sommerfeld, quoted in the paper cited in note 2.}

The second mistake is the promotion of space-time, from a 4-dimensional diagram that we each find an extremely useful conceptual device, into what Bohr called a ``real essence''.   My diagram enables me to represent events from my past experience, together with my possible conjectures, deductions, or expectations for events that are not in my past, or that escaped my direct attention.  By  identifying my abstract diagram with an objective reality, I fool myself into regarding the diagram as a 4-dimensional arena in which my life is lived. 

The events I experience are complex extended entities and the clocks I use to assign times to my experiences are extended macroscopic devices.  To represent my actual experiences as a collection of mathematical points in a continuous space-time is a brilliant strategic simplification, but we ought not to confuse a cartoon that concisely attempts to represent our experience, with the experience itself.

\section{IV. The Now of one person}

If I take my experience of Now as the reality it clearly is to me, and recognize that space-time is an abstract diagram that I use to organize such experiences, then the place of the Now in physics becomes obvious.  At any moment I can picture my past experience in my diagram as a continuous time-like curve that terminates in my  Now.  As my Now recedes into memory it ceases to be the real state of affairs,\ft{I believe this is what Huw Price objects to.}  and is replaced in my expanding diagram by subsequent Nows. 

The motion of my Now as it traces out my trajectory in my diagram reflects the evident fact that as my watch advances I acquire more experiences to record in that diagram.  My Now advances at one second of personal experience for each second that passes on my watch.\ft{One form of the argument that there can be no Now in physics is that the Now would have to move through time at a certain rate.   Since rates measure passage through time,  a mysterious second time seems needed, distinct from ordinary time.  
This is to confuse time with clocks.  Time is an abstraction that helps us to coordinate the readings of many different clocks.  
Clocks at the same set of events can be unambiguously compared with each other.   The two relevant clocks here are my mechanical (or electronic) watch and my internal biological rhythms. }  According to special relativity this means that a second of experience elapses for each second of proper time along my trajectory.   This connection between my ongoing experience and a geometric feature of my diagram is  just the diagrammatic representation of the fact that if asked ``What time is it now [i.e.~Now]?" I  look at my watch and report what I read. 

\section{V. The Now of many people}

The recognition that science is a tool we use to describe and organize ``the manifold aspects of our experience" clearly provides a place in physics for the Now of any single person, whose experience can be represented as a succession of local events along a single time-like trajectory. 
If there is a problem of the Now, it can only lie in relating the Nows of many different people.\ft{This, of course, is not a complication any solipsist would entertain.}     Here is where relativity enters the story.   Different people are generally in different places.   If there were a globally agreed upon simultaneity --- Newton's ``absolute, true, and mathematical time'' --- there would be no problem aligning the different Nows of different people, but Einstein taught us that there is no such thing.   
If two Now experiences of two people are space-like separated, then whether they belong to a single global Now   is entirely a matter of personal convention.    But since space-like separated events cannot influence one another, there are no consequences of two people having different conventions about whether such separate experiences share the same now.   

There is only one fundamental constraint on the Nows of two different people.  When  two people  are {\it together\/} at a {\it single event\/}, if that event happens to be Now for either one of them, then it must be Now for them both.   Like the Now of any single person, this is an obvious feature of human experience.      When you and I are interacting  face-to-face --- communicating --- it is  simply unimaginable\ft{I may underestimate  what people are capable of imagining.   I have no zombies in my {\it Weltanschauung\/}.}  that a live encounter for me could be only a memory for you, or vice versa.   This is as  basic a fact about the many perceiving subjects of science as the {\it Sum; ergo Nunc est.\/}  

This commonality of my Now and your Now {\it whenever\/} the two of us are together requires  our Nows to coincide at each of two consecutive meetings.   And indeed,   this is just what we find whenever we have a conversation,  move apart and then come back together and have another conversation.    But at the tiny relative speeds at which people have  moved apart and back together throughout human history, the possibly complicating effect of relativistic time dilation on the advances of the different individual Nows is utterly negligible compared with the psychological width (at least many milliseconds) of each of the private Now experiences.    We can, however entertain the question of whether our Nows would coincide when we came back together, regardless  of how rapidly we moved back and forth and regardless of how long the journey.   

As noted above, it is a basic fact of relativity that my personal time --- the progress of my Now --- keeps pace with my proper time --- the reading of my watch.\ft{If it did not I could sense my state of uniform motion, in violation of the principle of relativity.}  This is all that is needed.   Consider two twins.  At home together in an inertial frame, their Nows coincide.   Then Alice flies off to a nearby star at 4/5 of the speed of light, turns around, and flies back home to Bob at the same speed.    If Bob's watch has advanced ten years between their farewell and their reunion, then Alice's has advanced  six.   Because each of their Nows has advanced in step with the watch each is carrying, the moment of their reunion continues to be Now for both of them.

So to understand the Now the space-time-filling layers provided by the absolute, true, and mathematical time of Newton should be condensed into a filamentary structure of many multiply intersecting space-time trajectories in many different space-time diagrams of many different people, each having their own experiences and their own Nows. Whenever two or more people happen to be together at a single common event, their trajectories intersect in all of their diagrams.   Because the Now is one of the ``aspects of our experience'' and all the events that all of us experience are represented by points of such a filamentary structure, this is all we need for a complete physical description of all of our Nows and all of their interrelations.  

The content of the preceding three paragraphs refutes the misconception that physics has nothing whatever to say about the local Now at a single event.   Everything physics has to say about our Nows accords with our experience, and physics even predicts that our future experiences of the Now will continue to have these features even in a world of interstellar travel at relativistic speeds.

\section{VI.  Conclusion}

The explanatory power of the QBism of Fuchs and Schack is not limited to issues in quantum foundations.   The problem of the Now arises from the long-standing exclusion from classical physics of the experience of the perceiving subject, along with an inappropriate identification of the formalism of physics with the reality of the natural world. That there is such a thing as the present moment  is an undeniably real part of the experience of every one of us. The fact that we have a useful formalism that represents our experience and seems not to contain a Now does not mean that Now is an illusion.   It means that we must not identify the formalism with the experience that the formalism was constructed to describe.  

Physics accounts for the Nows of relativistically traveling twins coinciding at both their reunion and their departure.    This shows that there is indeed a place for the Now in science.   What was missing was not the Now.   What was missing was a recognition that the goal of science is to bring order and coherence to the experience of the person who uses it.   

There is thus no ``problem of the Now''.     But for now\ft{And also, of course, for Now.} this may be evident only to Cbists.

\section{{\it Acknowledgments.\/}}  My realization that QBism straightens out many problems developed during a decade of arguments and conversations, face-to-face and by email, with Chris Fuchs and Ruediger Schack.   Their patience, clarity of thought, and intellectual integrity were crucial for enlightening this slow and stubborn student.  An early draft of my essay benefitted from their critical remarks,   but although they are responsible for my general state of mind, they are not to blame for the flaws in this version.

I'm also indebted to Huw Price for reminding me of my 2009 talk at the Perimeter Institute.$^{11}$  When I listened to it  I was struck by its relevance to my recent thinking about  QBism. 

\bye